\documentclass[aps,preprint,showpacs]{revtex4}
\usepackage{amssymb}
\usepackage{amsmath}
\usepackage[pdftex]{graphicx}

\begin{document}

\title{Vortices on a superconducting nanoshell: phase diagram and dynamics}
\author{V.N.\ Gladilin$^{1,2}$, J.Tempere$^{1,3}$, I.F. Silvera$^{3}$, J.T.
Devreese$^{1}$, V.V. Moshchalkov$^{2}$}
\affiliation{$^{1}$TFVS, Universiteit Antwerpen, Groenenborgerlaan 171, 2020 Antwerpen,
Belgium\\
$^{2}$ INPAC, K.U.Leuven, Celestijnenlaan 200 D, B-3001 Leuven, Belgium.\\
$^{3}$ Lyman Laboratory of Physics, Harvard University, Cambridge MA 02138,
USA.}

\begin{abstract}
In superconductors, the search for special vortex states such as giant
vortices focuses on laterally confined or nanopatterned thin superconducting
films, disks, rings, or polygons. {We examine the possibility to realize
giant vortex states and states with non-uniform vorticity on a
superconducting spherical nanoshell, due to the interplay of the topology
and the applied magnetic field.} We derive the phase diagram and identify
where, as a function of the applied magnetic field, the shell thickness and
the shell radius, these different vortex phases occur. Moreover, the curved
geometry allows these states (or a vortex lattice) to coexist with a
Meissner state, on the same curved film. We have examined the dynamics of
the decay of giant vortices or states with non-uniform vorticity into a
vortex lattice, when the magnetic field is adapted so that a phase boundary
is crossed.
\end{abstract}

\pacs{PACS}
\date{\today}
\maketitle

\section{Introduction}

Quantized vortices are a quintessential property of superfluids and
superconductors. The energetically favored state when multiple quanta of
vorticity are present, is a lattice of singly quantized vortices. In
ultracold Fermi gases, the recent observation of such a vortex lattice
formed the `smoking gun' proof for superfluidity \cite{ZwierleinNAT435}. In
nanoscopic superconducting samples, controlling the vortex behavior is
essential for the development of new devices based on fluxon dynamics~\cite%
{Moshchalkov93}. The confinement of Cooper pairs to the length scales
comparable to the correlation length also offers the prospect to probe
fundamentally new phase topologies predicted by the theory, such as giant 
\cite{MoshchalkovGiant,MiskoPRL90} and {ring-like vortices}~\cite%
{StenuitPC332}. This has led to renewed experimental efforts to observe
giant vortex states, both in superconductors~\cite{Giantvortex,MiskoPRB64}
and in superfluid atomic gases\cite{GiantvortexBEC}.

In this contribution, we argue that superconducting spherical nanoshells
form a promising candidate for realizing giant vortex states, and for
engineering phase transitions between those states and a vortex lattice.
Moreover, we show that nanoshells allow the co-existence of a Meissner state
and a vortex state in equilibrium on one and the same superconducting film.
Nanoshells are hybrid nanostructures consisting of a dielectric core
(usually a silicon oxide nanograin), coated with a thin layer of metal\cite%
{Nanoshell}. When the metal in its bulk form is a superconductor, the
nanoshell below the critical temperature will also exhibit superconductivity
in the thin shell around the isolating core.

The superconducting order parameter in the nanoshell is well described by a
macroscopic wave function $\psi =\left\vert \psi \right\vert e^{i\varphi }$
that obeys the coupled time-dependent Ginzburg-Landau (TDGL) equations.
Vortices are characterized as topological defects in the phase $\varphi $
(requiring a vanishing gap $\left\vert \psi \right\vert $). {For thin shells}%
, the description is simplified in two important ways. Firstly, {when the
shell thickness is much smaller than the London penetration depth, the
magnetic field will be only weakly perturbed by the nanoshell}. Secondly,
when the shell is thinner than {the coherence length}, the order parameter $%
\psi $ will not vary substantially in the radial direction in the shell;
that is, $\psi $ will only depend on the spherical angles $\Omega =\{\theta
,\phi \}$. In the radial direction, $\psi $ will be constant in the shell,
and zero outside it. Note that confining $\psi $ to the shell leads to an
effective Ginzburg-Landau parameter $\kappa $ that differs from its bulk
value. In Section II we present the formalism, and in Sec. III the results,
for thin shells. When the shell thickness is increased and becomes
non-negligible with respect to the penetration depth, the magnetic field
will be more strongly perturbed, and the field gradients affect the
energetics. This case and the effect on the phase diagram are discussed in
Sec. IV. Finally, we summarize the results for vortices in nanoshells in
Sec. V.

\section{Ginzburg-Landau formalism on thin shells}

We assume that the shell is sufficiently thin for neglecting variations of
the order parameter across the shell. In other words, the order parameter $%
\psi $ will only depend on the spherical angles $\Omega =\{\theta ,\phi \}$.
We use the spherical coordinates $r$, $\theta $, $\phi $ with the origin at
the center of the sphere. The angle $\theta $ is counted from the $z$-axis
parallel to the external homogeneous magnetic field. Like in Ref.~%
\onlinecite{Zhao}, we will make the used variables dimensionless by
expressing lengths in units of $\sqrt{2}\lambda $, magnetic fields in units
of $\Phi _{0}/(4\pi \lambda ^{2})$, and the vector potential in units of $%
\Phi _{0}/(2\sqrt{2}\pi \lambda )$, where $\lambda $ is the penetration
depth, $\Phi _{0}=h/(2e)$ is the magnetic flux quantum, $h$ is the Planck
constant, and $e$ is the elementary charge. {Thus, the dimensionless
parameters $R$, $W$, and $H$ are linked to the radius of the nanoshell $%
\mathcal{R}$, its thickness $\mathcal{W}$, and the applied magnetic field $%
\mathcal{H}$ by the expressions $R=\mathcal{R}/(\sqrt{2}\lambda)$, $W=%
\mathcal{W}/(\sqrt{2}\lambda)$, and $H=4\pi \lambda ^{2}\mathcal{H}/\Phi
_{0} $, respectively.}

In our numerical treatment of superconducting states on spherical shells we
exploit the time-dependent Ginzburg-Landau equation, which is known to be a
powerful tool for studying both the dynamic and static properties of
superconductors. For a thin shell under consideration, the behavior of the
order parameter in a fixed (or slowly varying) magnetic field can be
described by the TDGL equation (cp.~\cite{hu72,kato91}) 
\begin{equation}
\frac{\partial \psi }{\partial \tau }=\left( \mathbf{\nabla }_{\Omega }-iR%
\mathbf{A}\right) ^{2}\psi +2(\kappa R)^{2}\psi (1-|\psi |^{2}),
\label{GL1dimless}
\end{equation}%
where $\kappa $ is the Ginzburg-Landau parameter, $\mathbf{A}$ is the
(dimensionless) vector potential, and $\mathbf{\nabla }_{\Omega }=\mathbf{e}%
_{\theta }(\partial /\partial \theta )+\mathbf{e}_{\phi }\sin ^{-1}(\theta
)(\partial /\partial \phi )$. The dimensionless variable $\tau $ is linked
to the time $t$ by the relation $\tau =Dt/\mathcal{R}^{2}$, with $D$, the
normal-state diffusion constant.

The vector potential $\mathbf{A}$ can be represented as a sum of the
contribution $\mathbf{A}_{1}$, related to supercurrents in the shell, and
the contribution $\mathbf{A}_{0}$, which corresponds to the external
magnetic field $\mathbf{H}$. The vector potential $\mathbf{A}_{0}$ is chosen
in the form 
\begin{equation}
\mathbf{A}_{0}=\mathbf{e}_{\phi }\frac{Hr\mathrm{sin}\theta }{2}.
\label{vecpot1c}
\end{equation}%
In the case of a constant applied magnetic field $\mathbf{H}$, with
increasing $\tau $ the function $\psi $, given by Eq.~(\ref{GL1dimless}),
approaches one of the (meta)stable states of the system (${\partial \psi }/{%
\partial \tau }\rightarrow 0$). The thermodynamically stable state is to be
found by comparing the Gibbs free energy for different solutions. The
difference in the Gibbs free energy between a superconducting state and the
normal state at the same magnetic field is given by the equation 
\begin{equation}
{\Delta G}=\frac{\Delta G_{0}}{4\pi \kappa ^{2}}\int\limits_{0}^{2\pi }d\phi
\int\limits_{0}^{\pi }d\theta \left[ \mathbf{A}_{1}\cdot \mathbf{j}-\kappa
^{2}|\psi |^{4}\right] ,
\end{equation}%
where $\Delta G_{0}$ corresponds to the superconducting state with no
vortices at $H=0$, i.e. the Meissner state present on the complete surface.
The dimensionless density of supercurrents is denoted by $\mathbf{j}$ and
expressed in units of $\Phi _{0}c/(8\sqrt{2}\pi ^{2}\lambda ^{3})$.

In this section, the shell is assumed to be sufficiently thin in order to
make negligible the magnetic fields, induced by supercurrents.
Correspondingly, we can neglect $\mathbf{A}_{1}$ as compared to $\mathbf{A}%
_{0}$. Then, as seen from Eqs.~(\ref{GL1dimless}) and (\ref{vecpot1c}), two
independent parameters, which govern the solution of Eq.~(\ref{GL1dimless}),
remain:

\begin{itemize}
\item the dimensionless size of the nanoshell $\rho\equiv \kappa R=\mathcal{R%
}/(\sqrt{2}\xi )$, determined by the ratio of the shell radius $\mathcal{R}$ 
{to the Ginzburg-Landau coherence length }$\xi $, and

\item the parameter $\eta \equiv HR^{2}/2=\pi \mathcal{H}\mathcal{R}%
^{2}/\Phi _{0}$, equal to the number of flux quanta of the applied field
that pass through the equatorial plane of the sphere.
\end{itemize}

\begin{figure}[tbp]
\centering \includegraphics[width=8.5cm]{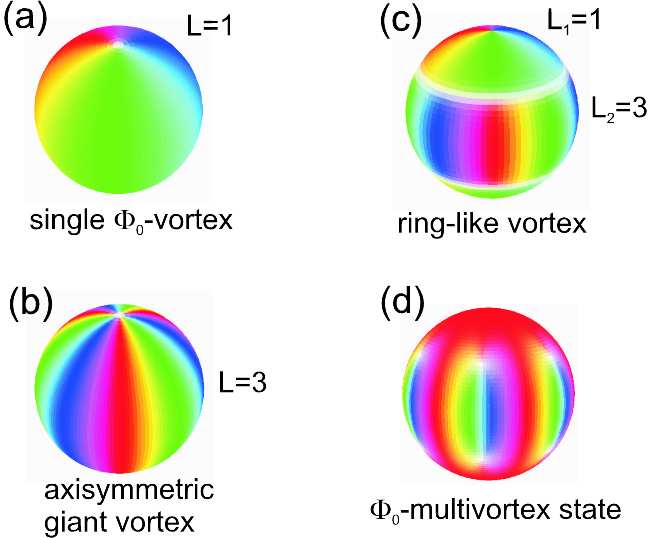}
\caption{A few examples of vortex structures on a spherical shell are
illustrated in this figure, to clarify the nomenclature used in the text.
The amplitude and phase of the order parameter are shown as the saturation
and hue of the color scale, respectively, in such a way that the vortex core
region is white. Panel (a) shows a single $\Phi _{0}$-vortex. In panel (b) a
giant vortex (carrying multiple quanta of flux) is depicted. Panel (c)
illustates a ring-like vortex separating regions with different angular
momentum. Finally, panel (d) shows the $\Phi _{0}$-multivortex state, where
an array of singly-quantized vortices is present.}
\label{fig0}
\end{figure}

When the magnetic field is increased beyond a critical value (computed
below), a first vortex appears for nanospheres with radius large enough to
sustain the vortex core, as depicted in Fig. \ref{fig0}, panel (a). Upon
further increasing the magnetic field, more quanta of flux can penetrates
the spherical surface. This can be accomodated in a variety of ways: for
example as a giant vortex carrying more than one quantum $\Phi _{0}$, shown
in panel (b) of Fig. \ref{fig0}. In this case, the angular momentum is
uniform over the spherical surface. It is also possible to envisage states
with non-uniform distributions of angular momentum: a value $L_{1}$ near the
poles and a value $L_{2}$ in a band near the equator. Such states are
characterized by a ring-like vortex separating the regions with different
angular momentum, as illustrated in panel (c) of Fig. \ref{fig0}. Also
states that do not have axial symmetry should be investigated: we will show
that these are in many case the stablest state and that they then consist of
an array of singly quantized vortices, illustrated in panel (d) of Fig. \ref%
{fig0}. Such states will be denoted by $\Phi _{0}$-multivortex states, to
emphasize that every vortex carries a single quantum of flux. In the next
subsections we start by investigating the axially symmetric states: giant
vortices and ring-like vortices. Then, in the next section we investigate
the condition under which those states decay into $\Phi _{0}$-multivortex
states$,$ and the dynamics of this decay.

\bigskip

\subsection{Giant vortex states}

\begin{figure}[tbp]
\centering \includegraphics[width=10.cm]{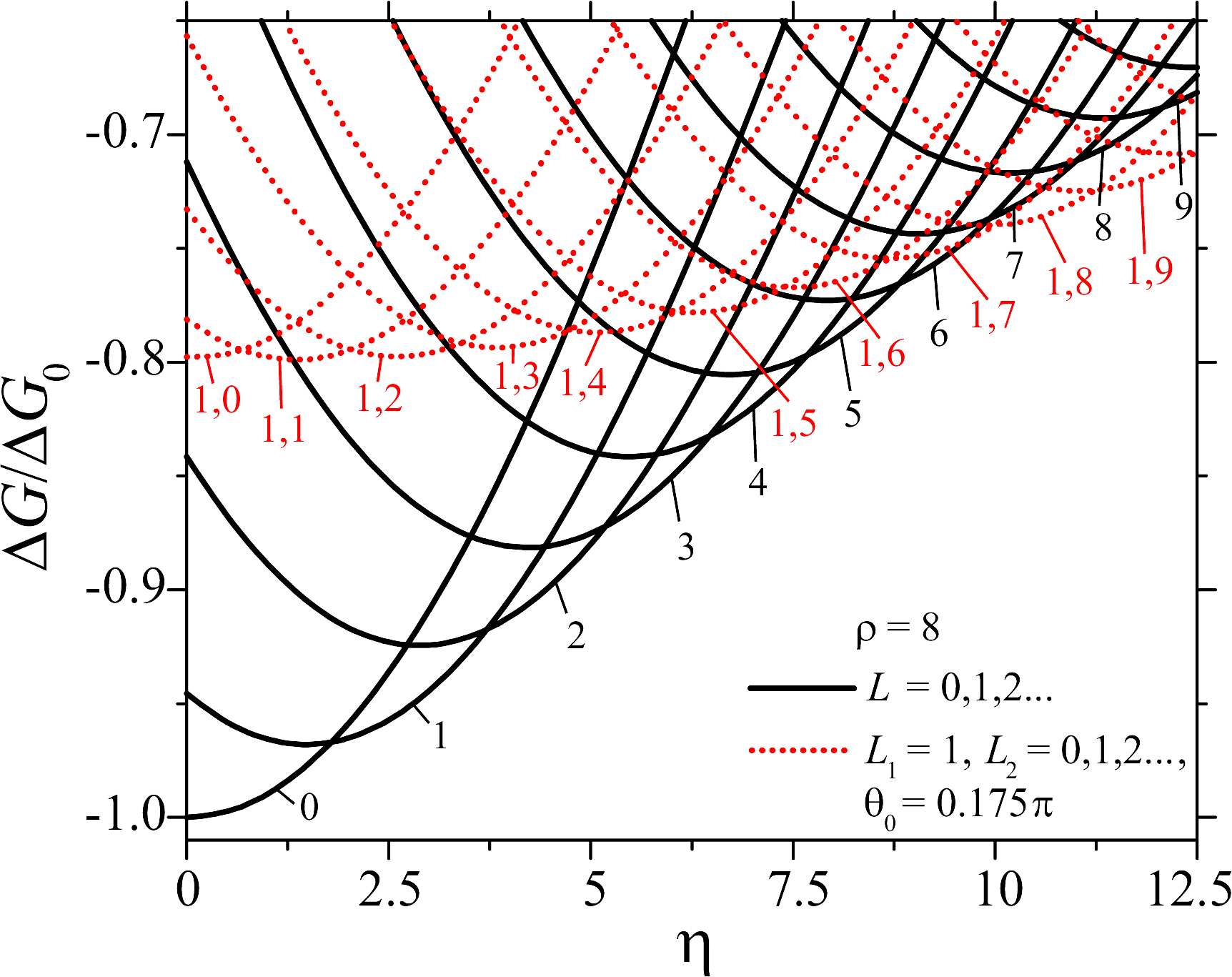}
\caption{The free energy difference is shown as a function of $\protect\eta $%
, the number of flux quanta that pass through the equatorial plane of the
sphere with radius $\protect\rho=8 $, for axially symmetric states. The
black solid curves are for states with uniform vorticity (and a giant
vortex). The red dotted lines are for states with ring-like vortices.}
\label{fig1}
\end{figure}

First, let us consider superconducting states, which keep the axial symmetry
of the system, so that the order parameter $\psi $ can be written in the
form $\psi =f(\theta )\mathrm{exp}(iL\phi )$. where $L$ has the sense of the
winding number (vorticity). Then for a stationary distribution $f(\theta )$
the Ginzburg-Landau equation~(\ref{GL1dimless}) reduces to the
one-dimensional equation 
\begin{equation}
\frac{\partial ^{2}f}{\partial \theta ^{2}}+\cot \theta \frac{\partial f}{%
\partial \theta }-\left( \frac{L}{\sin \theta }-{\eta \sin \theta }\right)
^{2}f+2\rho ^{2}f(1-f^{2})=0  \label{gl1}
\end{equation}%
with boundary conditions, determined by the requirement that the $\theta $%
-component of the current density must be zero at the $z$-axis: $\left. {%
\partial f}/{\partial \theta }\right\vert _{\theta =0,\pi }=0$. Solid lines
in Fig.~{\ref{fig1}} illustrate typical behavior of the free-energy
difference $\Delta G$ as a function of $\eta $ for cylindrically symmetric
states with different vorticity $L$. In the case of a thin spherical shell
with $\rho =8$, as illustrated in panel (a) of Fig. \ref{fig1}, the value of 
$L$ in the lowest cylindrically symmetric state increases with $\eta $ from
0 at $\eta =0$ to 9 at $\eta =12.5$. The modulus $f$ and phase $L\phi $ are
illustated in panel (b) of Fig. \ref{fig0}, using hue and saturation of the
color scale respectively. An increase of the applied magnetic field is seen
to result also in a significant increase of the Gibbs free energy of the
lowest state.

\subsection{Ring-like vortices}

In Ref.~\cite{Zhao}, when analyzing supercondicting states in hollow
cylinders, it was suggested that -- under certain conditions --
cylindrically symmetrical states with changing winding number can be more
energetically favorable than the states with uniform $L$. Our calculations
show that a similar situation occurs also in thin spherical shells with the
dimensionless size larger than $\rho \approx 6$ (i.e., for $\mathcal{R}%
\gtrsim 8.5\xi $), but as we will show in the next section, such states
decay into a lattice of singly-quantized vortices breaking the cylindrical
symmetry.

We have compared the Gibbs free energies for axially symmetric states with
uniform winding number $L$ and those for states where $L_{1}$, the winding
number at $0\leq \theta <\theta _{0}$ and $\pi -\theta _{0}<\theta \leq \pi $%
, differs from $L_{2}$, the winding number at $\theta _{0}<\theta <\pi
-\theta _{0}$. The order parameter of the latter states on the sphere are
illustrated in panel (c) of Fig. 1 -- we will refer to such states as
``ring-like vortex'' states. The continuity of the order parameter as a
function of $\theta $ requires vanishing $f(R,\theta )$ at the boundaries
between regions with different winding numbers, i.e., at $\theta =\theta
_{0} $ and $\theta =\pi -\theta _{0},$ as can be seen in panel (c) of Fig. %
\ref{fig0}. At sufficiently strong magnetic fields the Gibbs free energy for
states with ring-like vortices can become lower than that for states
characterized by a unique winding number $L$ over the whole $\theta $ range. 
{This is illustrated by Fig.~{\ref{fig1}}, where the dotted lines show the
calculated free-energy difference $\Delta G(\eta )$ for states with $L_{1}=1$
and different $L_{2}$ on a shell with $\rho =8$ in the case of $\theta
_{0}=0.175\pi $. It is worth mentioning that the values $\theta _{0}^{%
\mathrm{(min)}}$, which minimize $\Delta G$ for the lowest ring-like vortex
state at a given $\eta $, are rather insensitive to the dimensionless
nanoshell size $\rho $ (at least, for $\rho \leq 10$). At the same time, the
parameter $\theta _{0}^{\mathrm{(min)}}$ is an increasing function of $\eta $%
. Thus, our calculations show that this parameter changes from $\theta _{0}^{%
\mathrm{(min)}}\approx 0.12\pi $ to $\theta _{0}^{\mathrm{(min)}}\approx
0.2\pi $ when increasing $\eta $ from 7 to 15. However, moderate variations
of $\theta _{0}$ around $\theta _{0}^{\mathrm{(min)}}$ only slightly affect $%
\Delta G$ for the lowest state with ring-like vortices. That is why in Fig.~{%
\ref{fig1}} we restricted ourselves to the case of a fixed value $\theta
_{0}=0.175\pi $, which coincides with $\theta _{0}^{\mathrm{(min)}}$ at $%
\eta =10$. }

In Fig.~{\ref{fig1}}, the curve labeled with `1,1' corresponds to the state
with a ring-like vortex and uniform vorticity, which is qualitativley
similar to the states analyzed in Ref.~\onlinecite{StenuitPC332}. The free
energy of this state is always significantly higher than $\Delta G$ for the
lowest giant vortex states with point-like core. As can be further seen from
Fig.~{\ref{fig1}}, at $\eta \geq 10$ the ring-like vortex states ($L_{1}\neq
L_{2}$) appear the most energetically favorable among the axially symmetric
states. Thus, at $\eta \geq 12.5$, the difference in $\Delta G/\Delta G_{0}$
between the state with $L=9$ and the state with $L_{1}=1$, $L_{2}=9$ is
larger than 0.025. At even larger values of $\eta $, states with three
different regions of vorticity $L_{1},L_{2},L_{3}$ (characterized by two
ring-like vortices) can become stable. However, our calculations show that
in shells with $\eta \gtrsim 6$, where such ring-like vortices allow for
decreasing the free energy of giant vortex states as compared to the case of
a giant vortex, even lower values of $\Delta G$ can be achieved by breaking
up the ring like vortex (or vortices) into an array of singly-quantized
vortices. A natural question arises of how stable are the aforedescribed
giant vortex states with respect to decay into multiple singly quantized
vortices. In order to answer this question, one has to return to the TDGL
equation~(\ref{GL1dimless}).

\bigskip

\subsection{Numerical treatment}

The finite-difference scheme, applied here to solve Eq.~(\ref{GL1dimless}),
is similar to that of Ref.~\onlinecite{kato91}, with necessary adaptations
to the case of a spherical 2D-system. Two-dimensional grids, used in our
calculations, typically have $\gtrsim 100$ equally spaced nodes in the $%
\theta $-interval from 0 to $\pi $ and $\gtrsim 150$ equally spaced nodes in
the $\phi $-interval from 0 to $2\pi $. Cyclic boundary conditions for $\psi 
$ are applied at $\phi =0$ and $\phi =2\pi $. The boundary conditions at $%
\theta =0$ and $\theta =\pi $ are determined by the requirement $\left. \psi
\right\vert _{\theta =0,\pi }=\mathrm{const}(\phi )$. The step of the time
variable $\tau $ is automatically adapted in the course of calculation. This
adaptation is aimed to minimize the number of steps in $\tau $, necessary
for approaching a steady solution of Eq.~(\ref{GL1dimless}), and -- at the
same time -- to keep the solving procedure convergent. On average, the step
in $\tau $ is $\sim 10^{-5}$ to $\sim 10^{-4}$ depending on the used grid as
well as on $\rho $ and $\eta $. When starting at $\tau =0$ from a random
distribution of $\psi $ (with $|\psi |\ll 1$), a (meta)stable solution of
Eq.~(\ref{GL1dimless}) is achieved typically at $\tau \lesssim 100$. When
analyzing (meta)stability of states in a spherical shell, one has to keep in
mind that a transition between states with different vorticity, in general,
requires symmetry breaking. This means that simulations, which assume a
perfectly symmetric spherical nanoshell, would tend to overestimate
stability of a state with respect to a possible transition to another state
with lower free energy. In order to model the effect of imperfections,
inevitably present in realistic nanoshells, we consider spherical shells
with small angular variations $\delta \rho (\theta ,\phi )$ of the parameter 
$\rho $. Importantly, for relative magnitudes $|\delta \rho |/\rho $ ranging
roughly from $\sim 10^{-8}$ to $\sim 10^{-3}$, the results of simulations
practically do not depend on a specific choice of the magnitude and
distribution of these non-homogeneities. {An appreciable effect of those
imperfections on stable distributions of the order parameter appears only
for $|\delta \rho |/\rho >0.1$.}

\section{Results and discussion for thin shells}

\subsection{Decay of giant and ring-like vortices}

\begin{figure}[tbp]
\centering \includegraphics[width=8.3cm]{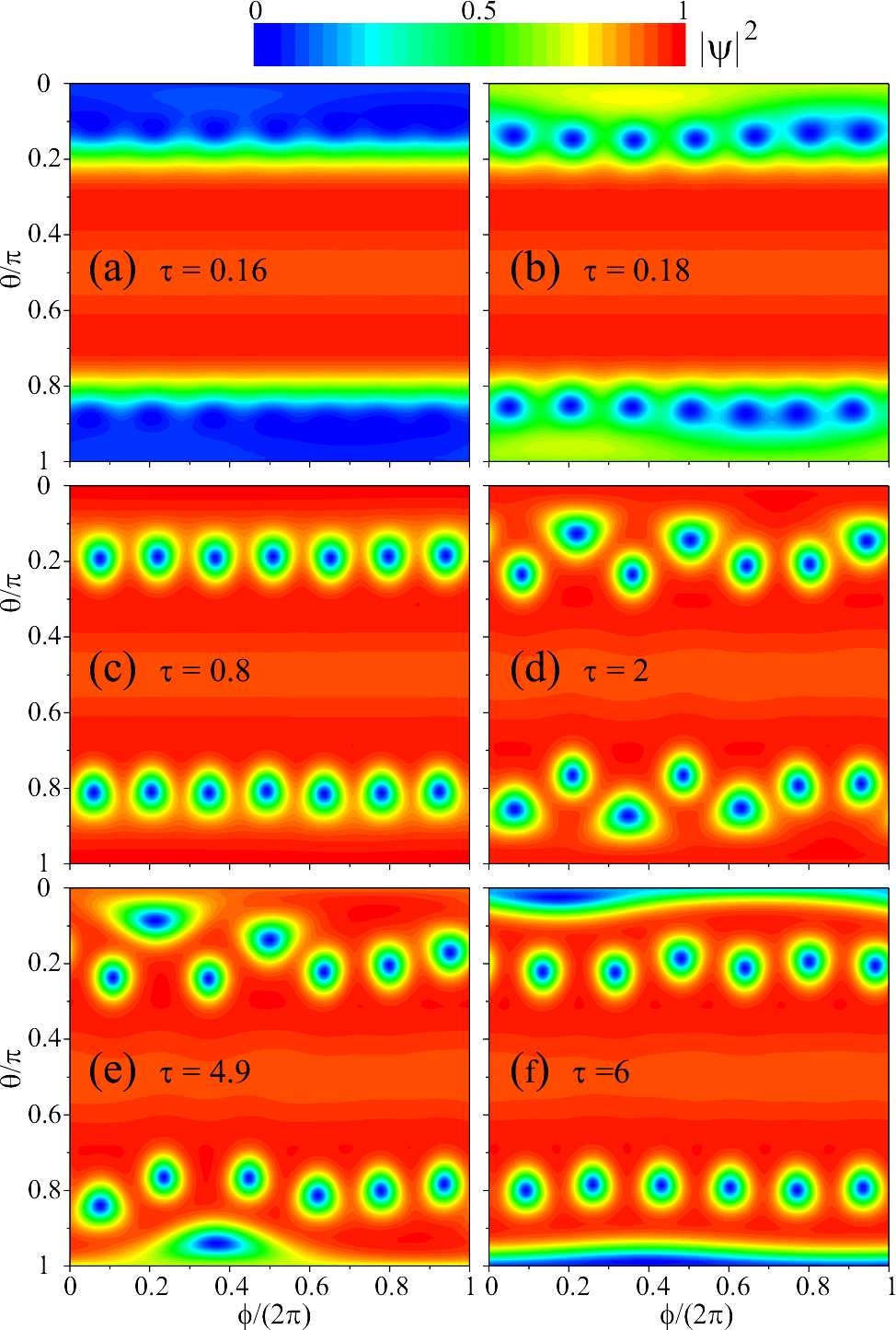}
\caption{ Evolution of the angular distribution of the squared modulus of
the order parameter in a thin spherical superconducting shell with $\protect%
\rho = 8$ at $\protect\eta=10$ in the case when the initial state (at $%
\protect\tau =0$) is a giant vortex with $L = 7$. Different panels
correspond to different time $\protect\tau$. }
\label{fig2}
\end{figure}

\begin{figure}[tbp]
\centering \includegraphics[width=8.3cm]{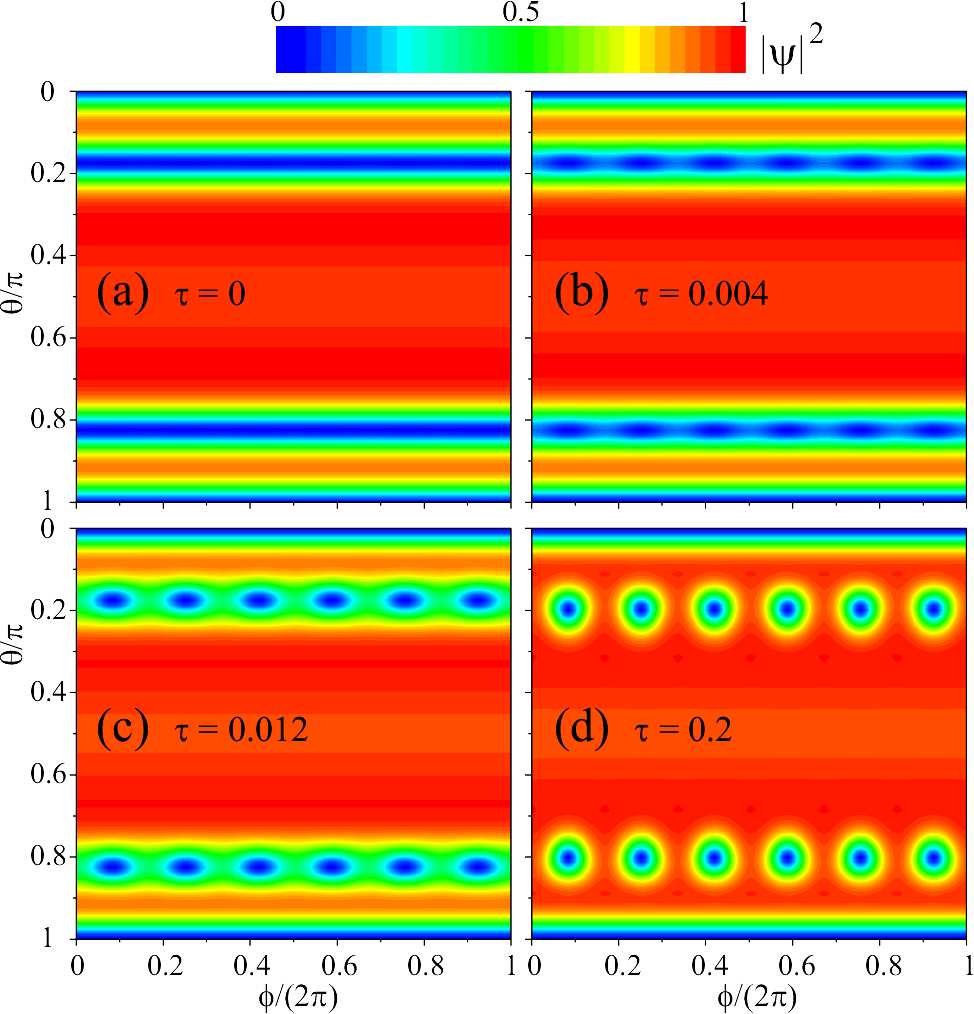}
\caption{ Evolution of the angular distribution of the squared modulus of
the order parameter in a thin spherical superconducting shell with $\protect%
\rho = 8$ at $\protect\eta=10$ in the case when the initial state (at $%
\protect\tau =0$) is a ring-like vortex state with $L_1=1$ for $0 \leq 
\protect\theta /\protect\pi < 0.175$ and $0.825 < \protect\theta /\protect%
\pi \leq 1$; $L_2 = 7$ for $0.175 \leq \protect\theta /\protect\pi \leq
0.825 $. Different panels correspond to different time $\protect\tau$. }
\label{fig3}
\end{figure}

In order to examine {the} stability of giant vortex states with respect to
decay into multivortex states, we apply {the computation scheme described in
the previous section}, starting at $\tau =0$ from a distribution of $\psi $ {%
that} corresponds to a giant or ring-like vortex state. Typical examples of
the evolution of the {order parameter distributions are shown in Figs.~\ref%
{fig2} and \ref{fig3} for the cases when the initial state is a giant vortex
and a ring-like vortex, respectively. In the case of $\rho =8$ and $\eta =10$%
, the thermodynamically stable state corresponds to 7 pairs of vortices with
a single quantum }$\Phi _{0}$ {of flux each, and it has a relative free
energy $\Delta G/\Delta G_{0}\approx -0.812${, approximately 0.07 lower than
the value of $\Delta G/\Delta G_{0}$ for the lowest giant vortex state (see
Fig.~\ref{fig1}). }Such lattices of vortices with each a singly flux quantum 
}$\Phi _{0}$ will be denoted as ``$\Phi _{0}$-multivortex states''.

As illustrated by Figs.~\ref{fig2}(a) to \ref{fig2}(c), within a $\tau $%
-interval $\sim 1$ the initial giant vortex state with $L=7$ transforms into
a chain of 7 singly quantized vortices, which surround each pole of the
sphere. In the course of the further rearrangement of the vortex pattern,
one of the vortices moves to the pole, while the remaining 6 vortices tend
to form a symmetric chain around the pole (see Figs.~\ref{fig2}(d) to \ref%
{fig2}(f)). A free-energy gain due to this rearrangement is by one order of
magnitude smaller than that due to the decay of the initial giant vortex
into single vortices. Correspondingly, the $\tau $-interval, necessary for
this rearrangement, appears to be relatively long: only at $\tau \approx 10$
the solution reaches the equilibrium symmetric configuration of vortices
(not shown in Fig.~\ref{fig2}), similar to that found in Ref.~%
\onlinecite{Du04}, where $\Phi _{0}$-multivortex {states} on a thin hollow
sphere were studied in detail. As seen from Fig.~\ref{fig3}, the transition
from a ring-like vortex state ($L_{1}=1$ for $0\leq \theta /\pi <0.175$ and $%
0.825<\theta /\pi \leq 1$; $L_{2}=7$ for $0.175\leq \theta /\pi \leq 0.825$)
to a $\Phi _{0}$-multivortex {state} is even faster. The ring-like vortex
core, which is present in the initial state [see Fig.~\ref{fig3}(a)], decays
into a chain of 6 single vortices very quickly: clear signatures of this
decay can be found already at $\tau <0.01$ [see Fig.~\ref{fig3}(b)]. The
equilibrium state with a vortex at the pole and 6 vortices, symmetrically
surrounding the pole, is formed already at $\tau =0.2$.

The results of our calculations clearly indicate that giant and ring-like
vortex states are rather unstable in spherical shells with relatively large $%
\rho $. This does not mean, however, that giant vortex states on a spherical
shell are never stable. A decrease of the shell radius and/or an increase of
the applied magnetic field enhance the role of the Lorenz forces, which act
on the supercurrents and tend to drive vortices towards the poles of the
shell. As a result, for sufficiently small $\rho $ and sufficiently large $%
\eta $, the distance between vortex cores in a $\Phi _{0}$-multivortex {state%
} becomes so small, that {physically} a $\Phi _{0}$-multivortex {state}
appears undistinguishable from the corresponding giant vortex state. A
similar continuous transition from a $\Phi _{0}$-multivortex {state} to a
giant vortex states with increasing magnetic field was recently found when
solving the linearized Ginzburg-Landau equation for superconducting
spherical {grains}~\cite{Baelus}. Of course, in the case of such a
continuous transition, the boundary between thermodynamically stable $\Phi
_{0}$-multivortex {states} and giant vortex states can be drawn only
approximately. As a criterion of a transition from a multivortex state to a {%
giant vortex} state, here we have chosen the condition that the angular
distance of vortex cores from the pole becomes smaller than $(10\rho )^{-1}$.

\subsection{Phase diagram for thin shells}

\begin{figure}[tbp]
\centering \includegraphics[width=8.3cm]{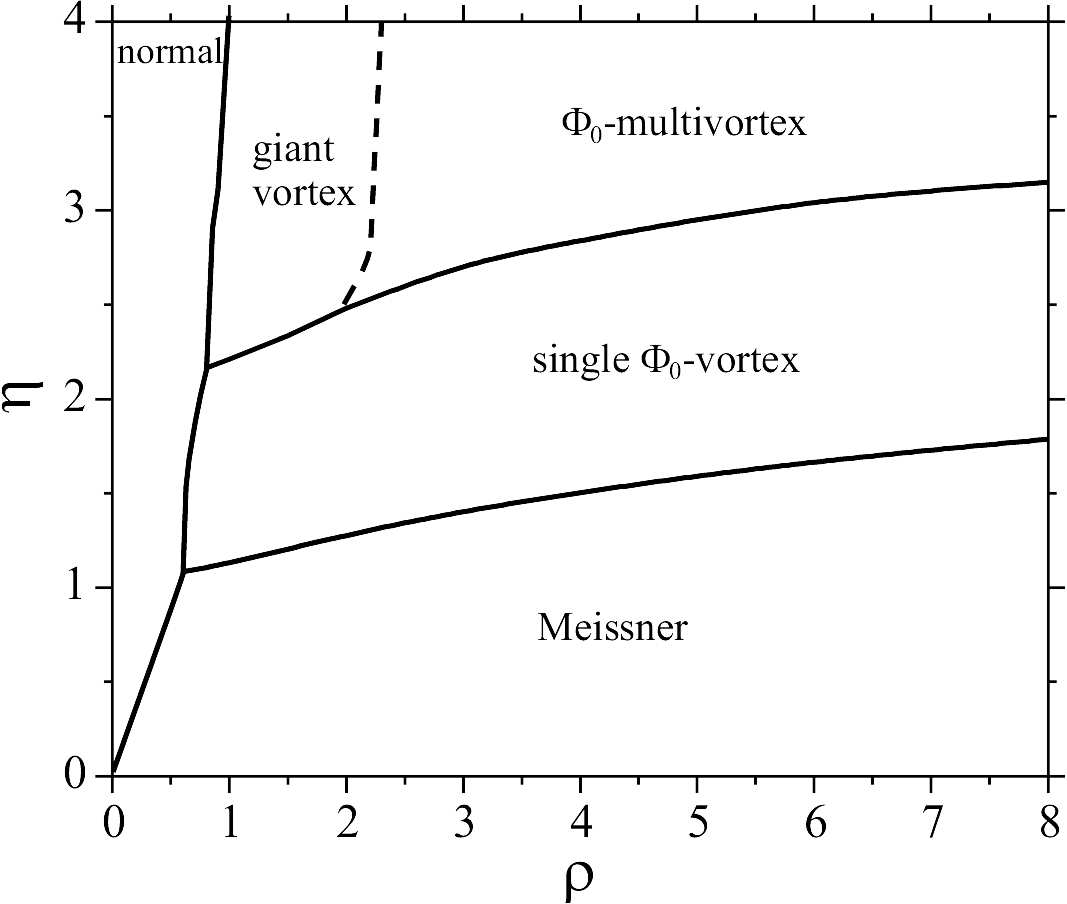}
\caption{ Phase diagram for thin spherical superconducting shells in the $(%
\protect\rho, \protect\eta)$-plane. The boundaries between the regions with
the thermodynamically stable normal state, the Meissner state, the single $%
\Phi_0$-vortex state, giant vortex state, and $\Phi_0$-multivortex states
are shown by solid lines. The dashed line approximately indicates the
boundary between the regions, where giant vortex states or $\Phi_0$%
-multivortex states are the thermodynamically stable state.}
\label{fig4}
\end{figure}

Our results, related to thermodynamically stable states on thin spherical
shells, are summarized in Fig.~\ref{fig4}, where the solid lines indicate
boundaries of stability regions for the normal state, the superconducting
Meissner states, single $\Phi _{0}$-vortex states, giant vortex and $\Phi
_{0}$-multivortex states. The dashed line indicates the boundary between the
regions, where giant vortex states (to the left from this line) or $\Phi
_{0} $-multivortex states (to the right from this line) are
thermodynamically stable. As seen from Fig.~\ref{fig4}, formation of
vortices can be energetically advantageous only on sufficiently large
shells: at $\rho \geq 0.63$ for states with $L=1$, at $\rho \geq 0.85$ for
giant vortex states, and at $\rho \geq 1.95$ for $\Phi _{0}$-multivortex
states.

\begin{figure}[tbp]
\centering \includegraphics[width=7.5cm]{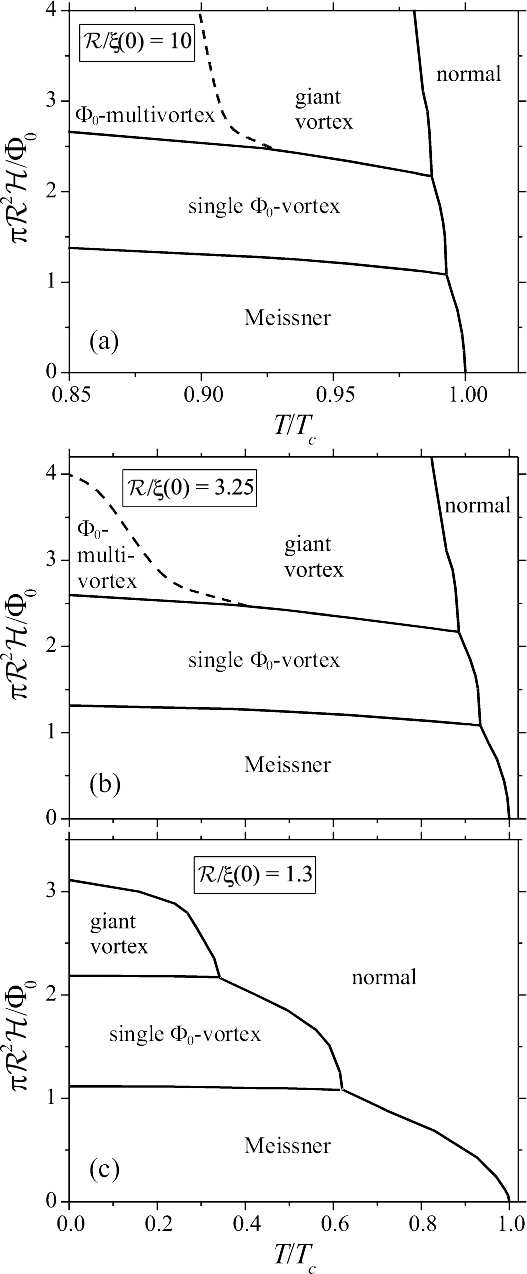}
\caption{Phase boundaries for thin spherical superconducting layers with
different radius $\mathcal{R}$ as a function of the temperature $T$ and the
applied magnetic field $\mathcal{H}$. Boundaries between the
thermodynamically stable normal state, Meissner state, single $\Phi_0$%
-vortex state, giant vortex state, and $\Phi_0$-multivortex states are shown
by solid lines. The dashed line approximately indicates the boundary between
the regions, where giant vortex states or $\Phi_0$-multivortex states are
the thermodynamically stable state.}
\label{fig5}
\end{figure}

While in Fig.~\ref{fig4} the phase diagram for thin spherical
superconducting shells is shown in the $(\rho ,\eta )$-plane, it seems
interesting to analyze the phase boundaries also in more common form: in
terms of the applied magnetic field $\mathcal{H}$ and the temperature $T$.
We assume that the temperature dependence of the penetration depth $\lambda $
is described by the empirical relation $\lambda (T)=\lambda (0)/\sqrt{%
1-(T/T_{c})^{4}}$, while the (less important) temperature dependence of the
Ginzburg-Landau parameter $\kappa $ is roughly given by the expression $%
\kappa (T)=\kappa (0)/[1+(T/T_{c})^{2}]$ (see, e.g., Ref.~%
\onlinecite{Tinkham}). In Fig.~\ref{fig5} we plot the phase boundaries for
thermodynamically stable normal states, Meissner states, single $\Phi _{0}$%
-vortex, giant vortex, and $\Phi _{0}$-multivortex states on spherical
superconducting shells with different radius $\mathcal{R}$, measured in
units of the zero-temperature Ginzburg-Landau coherence length $\xi (0)$. As
illustrated in Fig.~\ref{fig5}(a), in the case, where the radius $\mathcal{R}
$ is much larger than $\xi (0)$, giant vortex states are thermodynamically
stable (at moderate applied magnetic fields) only in the close vicinity of
the critical temperature $T_{c}$. With decreasing $\mathcal{R}$, the
stability range of giant vortex states gradually extends towards lower
temperatures and lower values of the magnetic flux through the shell.
Correspondingly, in sufficiently small shells the stability range of $\Phi
_{0}$-multivortex states is restricted to a relatively narrow interval of $%
\mathcal{H}$ and to $T$, significantly lower than $T_{c}$ [see Fig.~\ref%
{fig5}(b)]. On even smaller shells (with $\mathcal{R}$ close to $\xi (0)$)
no stable $\Phi _{0}$-multivortex states are possible [see Fig.~\ref{fig5}%
(c)]. For those small shells, the superconducting phase persists only for
relatively weak (few $\Phi _{0}$) magnetic fluxes through the shell. At the
same time, as implied by a comparison between panels (b) and (c) in Fig.~\ref%
{fig5}, the values of the applied magnetic field $\mathcal{H}$, which
correspond to transitions with an increase of vorticity by 1, become
significantly higher when decreasing the shell size.

\section{ Vortex states on thick shells}

\subsection{Magnetization effects in thick shells}

Now, let us extend our analysis to the case of relatively thick spherical
shells, where magnetic fields, induced by supercurrents, which flow in a
shell, are of non-negligible. At the same time, we assume that the thickness
of a shell is still sufficiently small for neglecting variations of the
order parameter $\psi$ and of the vector potential $\mathbf{A}=\mathbf{A}_0+%
\mathbf{A}_1$ across the layer. For such a shell, also currents across the
layer can be neglected. Expressing the vector potential $\mathbf{A}_1$
through the density of current $\mathbf{j}$ as 
\begin{equation}
\mathbf{A}_1(\mathbf{r})= \frac{1}{2\pi}\int d^3r^{\prime}\frac{\mathbf{j}(%
\mathbf{r}^{\prime})}{|\mathbf{r}-\mathbf{r}^{\prime}|}.  \label{vecpot12}
\end{equation}
the non-negligible components of the product $R\mathbf{A}_{1}$, which enters
Eq.~(\ref{GL1dimless}), can be written down in the following form: 
\begin{eqnarray}
R{A}_{1\theta} &=&\frac{WR}{2\sqrt{2}\pi} \int\limits_0^\pi d\theta^\prime
\sin\theta^\prime \int\limits_0^{2\pi} d\phi^\prime \left\{\left [\sin\theta 
\mathrm{sin}\theta^\prime +\cos\theta \cos\theta^\prime \cos \left(\phi
-\phi^\prime \right)\right] Rj_{\theta^\prime}\left(\theta^\prime,
\phi^\prime \right) \right.  \notag \\
&&\left. +\cos\theta\sin \left(\phi -\phi^\prime \right)
Rj_{\phi^\prime}\left(\theta^\prime, \phi^\prime \right)\right\} \left[%
1-\cos\theta \cos\theta^\prime-\sin\theta \sin\theta^\prime \cos \left(\phi
-\phi^\prime \right)\right]^{-1/2},  \label{atheta}
\end{eqnarray}
\begin{eqnarray}
R{A}_{1\phi} &=&\frac{WR}{2\sqrt{2}\pi} \int\limits_0^\pi d\theta^\prime
\sin\theta^\prime \int\limits_0^{2\pi} d\phi^\prime \left\{
\cos\theta^\prime \sin \left(\phi^\prime -\phi \right)
Rj_{\theta^\prime}\left(\theta^\prime, \phi^\prime \right) \right.  \notag \\
&&\left. +\cos \left(\phi -\phi^\prime \right)
Rj_{\phi^\prime}\left(\theta^\prime, \phi^\prime \right) \right\} \left[%
1-\cos\theta \cos\theta^\prime-\sin\theta \sin \theta^\prime \cos \left(\phi
-\phi^\prime \right)\right]^{-1/2},  \label{aphi}
\end{eqnarray}
where $W$ is the dimensionless thickness of the shell. On the other hand, in
the case of constant or slowly varying magnetic fields, using the relation 
\begin{equation}
\mathbf{j}= \mathrm{Re}\left[\psi^*\left(\frac{\bigtriangledown }{i}-\mathbf{%
A}\right)\psi\right],  \label{jj}
\end{equation}
the products $Rj_{\theta}$ and $Rj_{\phi}$, which enter Eqs.~(\ref{atheta})
and (\ref{aphi}), can be expressed through $\psi$, $R{A}_\theta$, and $R{A}%
_\phi$ as 
\begin{equation}
Rj_{\theta} = \mathrm{Im} \left[ \psi^* \frac{\partial \psi }{\partial \theta%
}\right] -R{A}_{1\theta} |\psi|^2,  \label{jtheta}
\end{equation}
\begin{equation}
Rj_{\phi}= \frac{1}{\sin\theta}\mathrm{Im}\left[\psi^*\frac{\partial \psi }{%
\partial \phi}\right]-\left(\eta\sin\theta +R{A}_{1\phi} \right)|\psi|^2.
\label{jphi}
\end{equation}
In order to find the order parameter $\psi$ and the corresponding vector
potential, we solve self-consistently the set of equations~(\ref{GL1dimless}%
), (\ref{atheta}), and (\ref{aphi}), using relations (\ref{jtheta}), and (%
\ref{jphi}). From Eqs.~(\ref{GL1dimless}), (\ref{atheta}), (\ref{aphi}), (%
\ref{jtheta}), and (\ref{jphi}), one can see that for relatively thick
shells under consideration a set of independent parameters, which govern the
solution, can be chosen as {$\eta$, $\rho$, and $\omega$, where the
introduced additional parameter $\omega\equiv WR=\mathcal{WR}/(2\lambda^2)$
is linearly proportional to the thickness of the nanoshell and to its radius.%
}

\begin{figure}[tbp]
\centering \includegraphics[width=16.4cm]{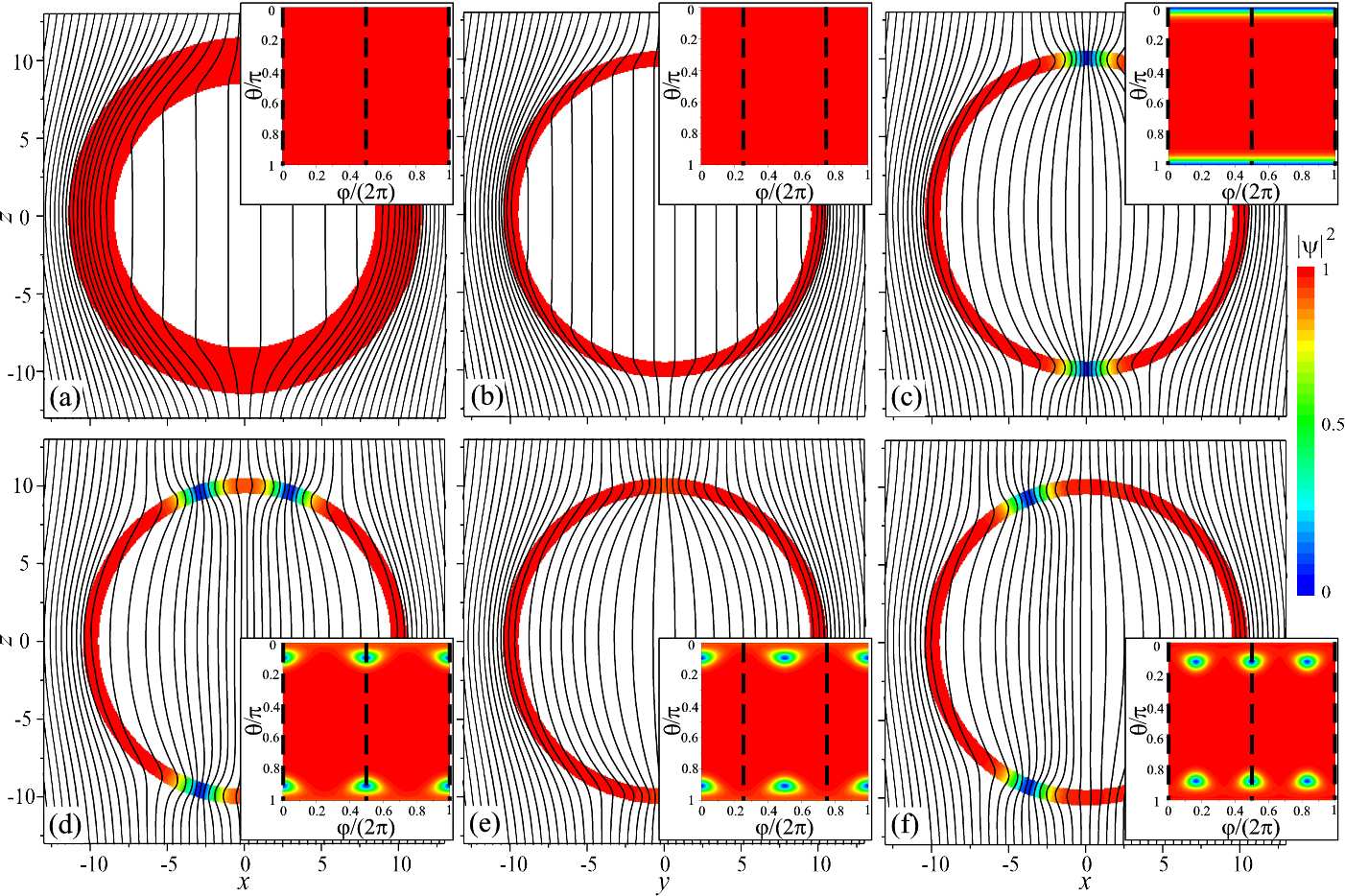}
\caption{Magnetic-field lines and distribution of the squared modulus of the
order parameter for thermodynamically stable states in superconducting
spherical shells with $\protect\kappa = 0.8$, $R=10$, $W=3$ [panel (a)] and $%
\protect\kappa = 0.8$, $R=10$, $W=1$ [panels (b) to (f)]. The results are
shown for the $xz$-cross-section [panels (a), (c), (d), (f)] and the $yz$%
-cross-section [panels (b), (e)] of the shell for different values of the
parameter $\protect\eta$ and vorticity $L$: $\protect\eta=5$, $L=0$ [panels
(a) and (b)], $\protect\eta=6.5$, $L=1$ [panel (c)], $\protect\eta=8$, $L=2$
[panels (d) and (e)], $\protect\eta=9.5$, $L=3$ [panel (f)]. Insets: angular
distributions of the squared modulus of the order parameter for the same
values of $\protect\eta$ and other parameters. Vertical dashed lines on each
inset correspond to the cross-section displayed on the main panel. }
\label{fig6}
\end{figure}

Figure~\ref{fig6} gives few examples of magnetic-field distributions, which
correspond to thermodynamically stable states in spherical shells with $\rho
=8$, $\omega =30$ and $\rho =8$, $\omega =10$. The patterns of
magnetic-field lines, displayed in Fig.~\ref{fig6}, are plotted for the
particular case of the Ginzburg-Landau parameter $\kappa =0.8$, the (mean)
dimensionless radius of the shell $R=10$, and the dimensionless thickness $%
W=3$ and $W=1$. In general, none of the three mutually orthogonal components
of the magnetic field $\mathbf{B}=\mathbf{\nabla }\times \mathbf{A}$ is
zero, so that the field lines are \textquotedblleft
three-dimensional\textquotedblright . In Fig.~\ref{fig6}, however, we
restrict ourselves to field-line patterns within symmetry planes, where the
field lines are \textquotedblleft flat\textquotedblright . As seen from Fig.~%
\ref{fig6}(a), even in the case of a relatively thick shell ($W=3$) the
magnetic fields, induced by the supercurrents in the Meissner state, are not
sufficient for complete screening of the applied magnetic field inside the
shell. Nevertheless, not only in the case of $W=3$ but also for a
significantly thinner shell with $W=1$ [Fig.~\ref{fig6}(b)], the net field
inside the shell is much weaker than $H$. In the case of the state with $L=1$%
, the magnetic flux, captured by a vortex pair, is seen as an increased
density of field lines at the poles of the sphere [Fig.~\ref{fig6}(c)]. At
the same time, in the depth of the sphere the magnetic field is relatively
homogeneous, only slightly increasing towards the $z$-axis. Also for states
with higher vorticity, a considerable local increase of the magnetic-flux
density takes place only at the vortex cores within the superconducting
shell, while in the depth of the sphere the density of magnetic-field lines
is considerably more homogeneous [see panels (d) to (f) in Fig.~\ref{fig6}].

\subsection{Phase diagram for thick shells}

\begin{figure}[tbp]
\centering \includegraphics[width=8.3cm]{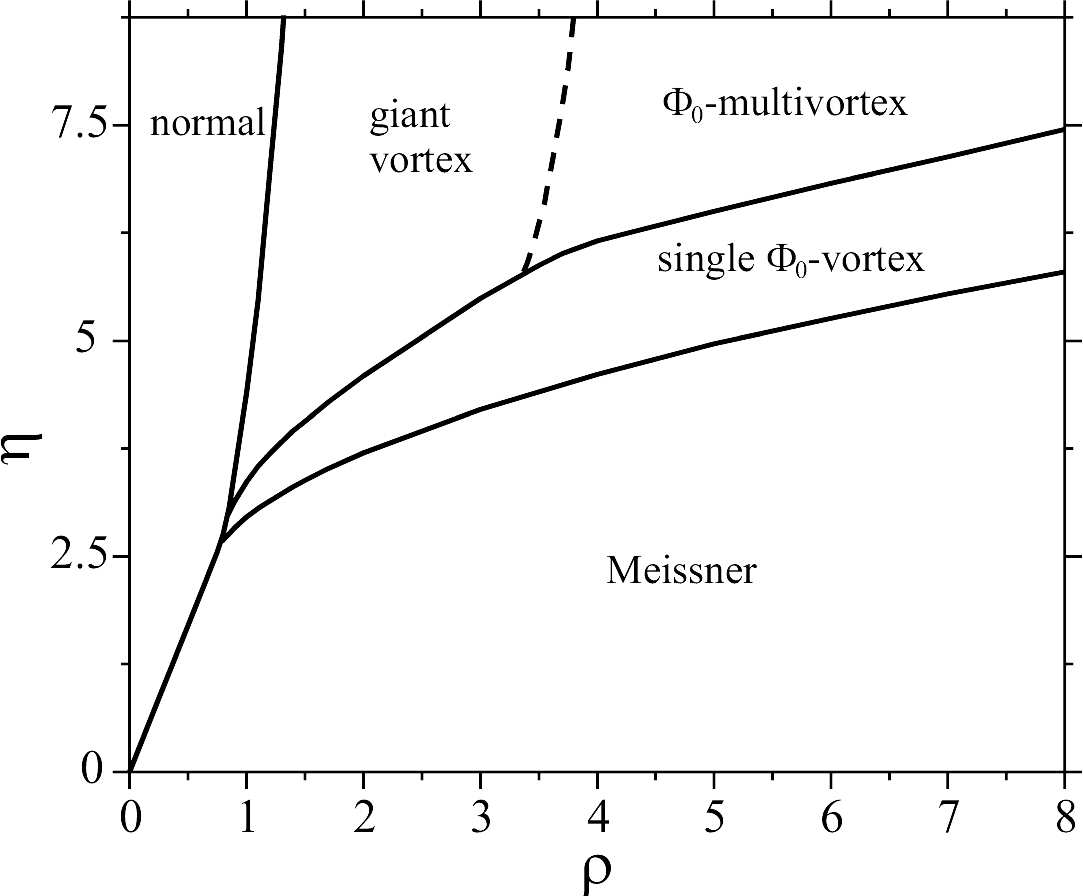}
\caption{ Phase diagram for thick ($\protect\omega=10$) spherical
superconducting shells in the $(\protect\rho, \protect\eta)$-plane. The
boundaries between the thermodynamically stable normal state, Meissner
state, single $\Phi_0$-vortex state, giant vortex state, and $\Phi_0$%
-multivortex states are shown by solid lines. The dashed line approximately
indicates the boundary between the regions, where giant vortex states or $%
\Phi_0$-multivortex states are the thermodynamically stable state.}
\label{fig7}
\end{figure}

As seen from Fig.~\ref{fig6}, the magnetic fields, induced by supercurrents,
can be considerably large even for shells with quite moderate thickness ($%
W\sim 1$). These fields strongly affect the stability range for
superconducting states with different vorticity in a spherical shell. In
Fig.~\ref{fig7}, we present the calculated phase diagram for relatively
thick spherical shells with $\omega =10$. As follows from a comparison of
Fig.~ \ref{fig7} to Fig.~\ref{fig4}, an increase of the thickness of a
spherical shell results in a well-pronounced shift of the boundaries between
states with different vorticity towards higher magnetic fields $\eta $. In
particular, for $\rho >0.8$, the range of $\eta $, where Meissner states are
thermodynamically stable, is more than two times wider in the case of $%
\omega =10$ as compared to the case of $\omega \rightarrow 0$. One can also
see that for a relatively thick spherical shell ($\omega =10$) the boundary
between giant vortex and $\Phi _{0}$-multivortex states is shifted towards
significantly larger values of $\rho $ as compared to those in the case of $%
\omega \rightarrow 0$. The increased stability of giant vortex states agree
with the results recently obtained by Baelus \emph{et al}. \cite{Baelus} for
the limit $W\rightarrow R$ of a full sphere, in the framework of linearized
Ginzburg-Landau equations.

\begin{figure}[tbp]
\centering \includegraphics[width=7.5cm]{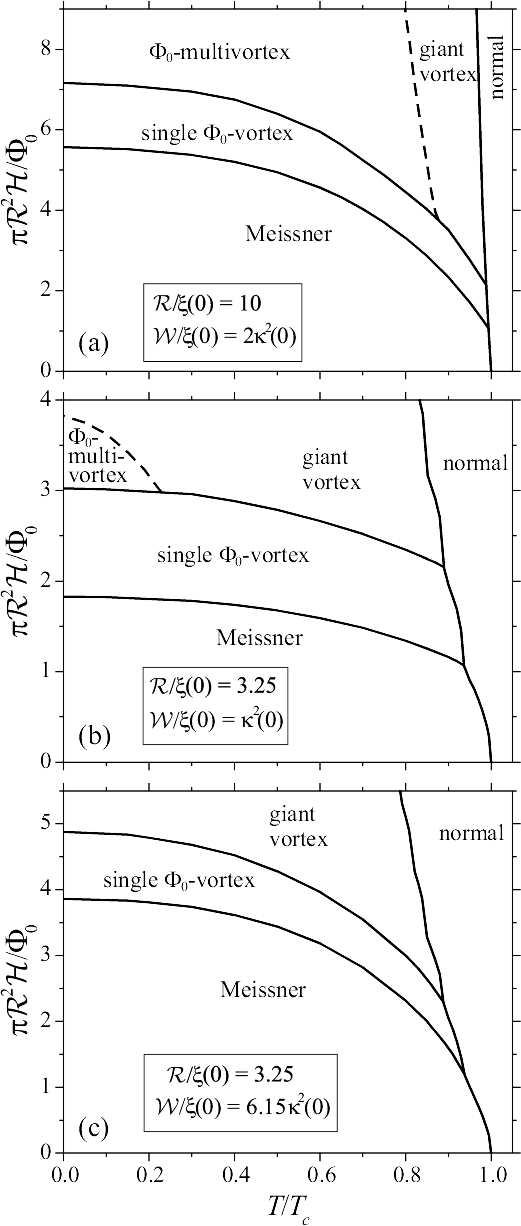}
\caption{Phase boundaries for spherical superconducting layers with
different radius $\mathcal{R}$ and thickness $\mathcal{W}$ as a function of
the temperature $T$ and the applied magnetic field $\mathcal{H}$. The
boundaries between the thermodynamically stable normal state, Meissner
state, single $\Phi_0$-vortex state, giant vortex state, and $\Phi_0$%
-multivortex states are shown by solid lines. The dashed line approximately
indicates the boundary between the regions, where giant vortex states or $%
\Phi_0$-multivortex states are the thermodynamically stable state.}
\label{fig8}
\end{figure}

In Fig.~\ref{fig8} the phase boundaries for thermodynamically stable normal
states, Meissner states, single $\Phi _{0}$-vortex, giant vortex, and $\Phi
_{0}$-multivortex state states are plotted in the $(T,\mathcal{H})$-plane.
The results are shown for shells with different radius $\mathcal{R}$ and
thickness $\mathcal{W}$. In order to keep the plots more universal, it is
convenient to express the thickness in units of $\xi (0)\kappa ^{2}(0)$. For
thick nanoshells, there is a much more pronounced increase of the transition
fields, which correspond to a change of vorticity, with lowering temperature
[cp. Figs.~\ref{fig8}(a) and~\ref{fig8}(c) to Figs.~\ref{fig5}(a) and~\ref%
{fig8}(c)]. When comparing Fig.~\ref{fig8}(a) to Fig.~\ref{fig5}(a) one can
also see that with increasing the nanoshell thickness the temperature range,
where giant vortices are thermodynamically stable, extends towards lower
temperatures. With decreasing the shell radius, this effect becomes quite
pronounced even for relatively small values of $\mathcal{W}/[\xi (0)\kappa
^{2}(0)]$ [cp. Fig.~\ref{fig8}(b) to Fig.~\ref{fig5}(b)]. In sufficiently
thick nanoshells, the temperature range, where $\Phi _{0}$-multivortex
states are thermodynamically stable, reduces to zero [see Fig.~\ref{fig8}%
(c)], although in thin nanoshells of the same radius this range is
relatively wide [see Fig.~\ref{fig5}(b)]. Of course, when the temperature
approaches $T_{c}$, the phase boundaries become almost insensitive to the
value of $\mathcal{W}$. Indeed, at $T\rightarrow T_{c}$ the parameter $%
\omega $ always goes to zero [due to an increase of the penetration depth $%
\lambda (T)$], so that any nanoshell appears effectively thin. 

\section{Conclusions}

Curving a superconducting film into a spherical shell changes its
vortex-related properties drastically due to topological constraints. The
hairy-sphere theorem~\cite{Brouwer} is a straightforward example of such a
constraint: it states that, in contrast to the situation on a flat film,
there exists no nonvanishing continuous tangent vector field on the sphere.
So, every nonvanishing supercurrent velocity field requires discontinuities,
such as vortices. The interplay between the Lorentz force due to an applied
field and the vortex superflow will force these vortices away from the
equator (leaving an equatorial \textquotedblleft Meissner
band\textquotedblright ) and towards the poles. This results in a `polar
trapping potential', which is nearly quadratic near the poles. When vortices
conglomerate at the poles, they may coalesce to form giant or ring-like
vortices, and these dynamics and phases are the topic of the present paper.

Three contributions to the energy should be kept in mind to interpret the
phase diagrams obtained in our calculations. First, to create a vortex, the
kinetic energy of the associated supercurrent (on the 2D spherical surface)
should be taken into account. This contribution increases when two vortices
with parallel vorticity are placed near each other, so it acts as a
repulsion between the vortices. Thus, it tends to favor splitting of the
giant vortices. Second, to create a vortex, the order parameter needs to be
suppressed over a region typically of the size of the {coherence} length.
The energy cost associated with this turns out to favor a multiply quantized
(giant) vortex over the corresponding $\Phi _{0}$-multivortex state. The
energy cost is relatively larger for a smaller sphere, since proportionally
a larger fraction of the total order parameter needs to be suppressed. The
balance between these two energy contributions can be used to qualitatively
understand the phase diagrams that we calculate for thin shells. Indeed, for
magnetic fields corresponding to multiple quanta of vorticity, the smaller
spheres will favor giant vortices, whereas the larger spheres favor the $%
\Phi _{0}$-multivortex state. Note that this contribution to the energy
strongly disfavors ring-like vortex states.

The third contribution to the energy is related to the gradients in the
magnetic field. When the shell is much thinner than the penetration depth,
the currents on the shell will not substantially perturb the applied field,
and this contribution plays no role. However, for thicker shells, this
contribution does become important -- as can be seen from Fig.~\ref{fig6},
the magnetic field is substantially perturbed. When a $\Phi _{0}$%
-multivortex lattice is present, the magnetic field flux is concentrated
near each vortex core, and shielded in between, leading to a larger magnetic
contribution to the energy than for a giant vortex. Thus, for a thick shell,
this contribution will favor the giant vortex state. This agrees with our
phase diagram showing that the region, where the giant vortex is stable, is
growing for thicker shells. 

The temperature dependence of the phase diagrams was studied
straightforwardly by taking temperature into account through the
Ginzburg-Landau parameters. When multiple quanta of vorticity are present,
we find that the giant vortex phase forms the preferred high-temperature
phase. This offers the prospect of probing a temperature-driven transition
between a giant vortex and a $\Phi _{0}$-multivortex state, alongside with a
magnetic-field driven transition. Moreover, the vortex dynamics are shown to
be not sensitive to {moderate} imperfections in the shell; the energy
contributions discussed here can overcome the pinning potential due to for
example thickness inhomogeneities -- such pinning potentials have in past
experimental work hampered the detection of the giant vortex state. This
robustness, together with the tunability of the phase diagram through a
limited set of controllable parameters, makes superconducting nanoshells
uniquely suited for the study of novel vortex states.

\begin{acknowledgments}
This work was supported by the Fund for Scientific Research - Flanders
projects Nos. G.0356.06, G.0115.06, G.0435.03, G.0306.00, the W.O.G. project
WO.025.99N, the GOA BOF UA 2000 UA ...
\end{acknowledgments}


\begin{thebibliography}{99}
\bibitem{ZwierleinNAT435} M. W. Zwierlein, J. R. Abo-Shaeer, A. Schirotzek,
C. H. Schunck and W. Ketterle, Nature \textbf{435}, 1047-1051 (23 June 2005).

\bibitem{Moshchalkov93} V.V. Moshchalkov, L. Gielen, M. Dhall\'{e}, C. Van
Haesendonck, Y. Bruynseraede, Nature \textbf{361}, 617 (1993).

\bibitem{MoshchalkovGiant} V. V. Moshchalkov, X. G. Qiu, and V. Bruyndoncx,
Phys. Rev. B \textbf{55}, 11793 (1997).

\bibitem{MiskoPRL90} V.R. Misko, V.M. Fomin, J.T. Devreese, V.V.
Moshchalkov, Phys. Rev. Lett. \textbf{90}, 147003 (2003).

\bibitem{StenuitPC332} G. Stenuit, J. Govaerts, D. Bertrand and O. van der
Aa, Physica C \textbf{332}, 277 (2000); ibid. Phys. Lett. A \textbf{267}, 56
(2000).

\bibitem{Giantvortex} V. Bruyndoncx, J. G. Rodrigo, T. Puig, L. Van Look,
and V. V. Moshchalkov, Phys. Rev. B \textbf{60}, 4285 - 4292 (1999); R.
Jonckheere \emph{et al.}, Phys. Rev. Lett. \textbf{85}, 1528 (2000); ibid.
Phys. Rev. Lett. \textbf{86}, 1663 (2001); A. Kanda \emph{et al.}, Phys.
Rev. Lett. \textbf{93}, 257002 (2004).

\bibitem{MiskoPRB64} V.R. Misko, V.M. Fomin and J.T. Devreese, Phys. Rev. B 
\textbf{64} (2001).

\bibitem{GiantvortexBEC} U.R. Fischer and G. Baym, Phys. Rev. Lett. \textbf{%
90}, 140402 (2003); P. Engels, I. Coddington, P. C. Haljan, V. Schweikhard,
and E. A. Cornell, Phys. Rev. Lett. \textbf{90}, 170405 (2003).

\bibitem{Nanoshell} R.D. Averitt, D. Sarkar, and N.J. Halas, Phys.\ Rev.
Lett. \textbf{78}, 4217 (1997); S.J. Oldenburg, R.D. Averitt, S.L. Westcott,
and N.J. Halas, Chem. Phys. Lett. \textbf{288}, 243 (1998).

\bibitem{hu72} C. R. Hu, R. S. Thomson, Phys. Rev. B \textbf{6}, 110 (1972).

\bibitem{kato91} R. Kato, Y. Enomoto, S. Maekawa, Phys. Rev. B \textbf{44},
6916 (1991).

\bibitem{Zhao} Hu Zhao, V. M. Fomin, J. T. Devreese, V. V. Moshchalkov,
Solid State Commun. \textbf{125}, 59 (2003).

\bibitem{Du04} Q. Du, L. Ju, J. Comp. Phys. \textbf{201}, 511 (2004); ibid.
Math. Comp. \textbf{74}, 1257 (2004).

\bibitem{Baelus} B. J. Baelus, D. Sun, F. M. Peeters, Phys. Rev. B \textbf{75%
}, 174523 (2007).

\bibitem{Tinkham} M. Tinkham, \textit{Introduction to Superconductivity}
(2nd ed., McGraw-Hill, New York, 1996).

\bibitem{Brouwer} L. E. J. Brouwer, Mathematische Annalen \textbf{71}, 97
(1912).
\end{thebibliography}
\end{document}